\documentclass{aa}

\usepackage[para]{threeparttable}
\usepackage{booktabs,caption}
\usepackage{graphicx}
\usepackage{hyperref}
\usepackage{soul}
\usepackage{color}
\usepackage{placeins}
\usepackage{csvsimple}
\usepackage{tabularx}
\usepackage{pdflscape}
\usepackage{afterpage}
\usepackage{txfonts}

\newcommand{\Teff}{\,T_\mathrm{eff}} 

\newcommand{\Ks}{\,K_\mathrm{S}} 

\begin{document}

   \title{A Catalogue of Solar-Like Oscillators Observed by TESS in 120-second and 20-second Cadence}

\author{Emily Hatt
          \inst{1}\thanks{E-mail: EXH698@student.bham.ac.uk}
          \and
          Martin B. Nielsen\inst{1, 2, 3}
          \and
          William J. Chaplin\inst{1, 2}
          \and 
          Warrick H. Ball \inst{1, 2}
          \and 
          Guy R. Davies\inst{1}
          \and 
          Timothy R. Bedding\inst{2, 4}
          \and 
          Derek L. Buzasi\inst{5}
          \and 
          Ashley Chontos\inst{6}
          \and
           Daniel Huber\inst{6}
          \and
          Cenk Kayhan\inst{7}
          \and
          Yaguang Li\inst{2, 4}
          \and
          Timothy R. White\inst{2, 4, 8}
          \and
          Chen Cheng \inst{4}
          \and
          Travis S. Metcalfe \inst{9}
          \and
          Dennis Stello \inst{10}}

      \institute{School of Physics and Astronomy, University of Birmingham, Birmingham B15 2TT, UK\\
              \email{exh698@student.bham.ac.uk}
              \and
              Stellar Astrophysics Centre (SAC), Department of Physics and Astronomy, Aarhus University, Ny
Munkegade 120, DK-8000 Aarhus C, Denmark
             \and
             Center for Space Science, NYUAD Institute, New York University Abu Dhabi, PO Box 129188, Abu
Dhabi, United Arab Emirates
             \and
             Sydney Institute for Astronomy (SIfA), School of Physics, University of Sydney, Camperdown, NSW 2006, Australia.
             \and
             Department of Chemistry and Physics, Florida Gulf Coast University, Fort Myers, FL 33965
             \and
             Institute for Astronomy, University of Hawai`i, 2680 Woodlawn Drive, Honolulu, HI 96822, USA
             \and 
             Department of Astronomy and Space Sciences, Science Faculty, Erciyes University, 38030 Melikgazi, Kayseri, Turkey
             \and 
             Research School of Astronomy and Astrophysics, Mount Stromlo Observatory, The Australian National University, Canberra, Australian Capital Territory, Australia.
             \and 
             White Dwarf Research Corporation, 9020 Brumm Trail, Golden, CO 80403, USA
             \and
             School of Physics, University of NSW, 2052, Australia}
   \date{}

   \date{}

  \abstract
   {The Transiting Exoplanet Survey Satellite (TESS) mission has provided photometric light curves for stars across nearly the entire sky. This allows for the application of asteroseismology to a pool of potential solar-like oscillators that is unprecedented in size.
}
   {We aim to produce a catalogue of solar-like oscillators observed by TESS in the 120-second and 20-second cadence modes. The catalogue is intended to highlight stars oscillating at frequencies above the TESS 30-minute cadence Nyquist frequency with the purpose of encompassing the main sequence and sub-giant evolutionary phases.  We aim to provide estimates for the global asteroseismic parameters $\nu_{\mathrm{max}}$ and $\Delta \nu$.}
   {We apply a new probabilistic detection algorithm to the 120-second and 20-second light curves of over 250,000 stars. This algorithm flags targets that show characteristic signatures of solar-like oscillations. We manually vet the resulting list of targets to confirm the presence of solar-like oscillations. Using the probability densities computed by the algorithm, we measure the  global asteroseismic parameters $\nu_{\mathrm{max}}$ and $\Delta \nu$.}
   {We produce a catalogue of 4,177 solar-like oscillators, reporting $\Delta \nu$ and $\nu_{\mathrm{max}}$ for 98\% of the total star count. The asteroseismic data reveals vast coverage of the HR diagram, populating the red giant branch, the subgiant regime and extending toward the main-sequence.}
   {A crossmatch with external catalogs shows that 25 of the detected solar-like oscillators are a component of a spectroscopic binary, and 28 are confirmed planet host stars. These results provide the potential for precise, independent asteroseismic constraints on these and any additional TESS targets of interest.}


   \maketitle
%
\section{Introduction}
Asteroseismology, the study of the intrinsic oscillations of stars, has revealed the physical properties of thousands of stars to high precision \citep[e.g.][]{2014ApJS..214...27M, 2014A&A...569A..21L, 2015A&A...580A.141L, 2017ApJS..233...23S,Yu_2018, 2019MNRAS.489.1753Y}. Solar-like oscillators, where-in modes are excited and damped by the turbulent motion of gas in the outer convection zone, have been of particular interest due to the host of identifiable overtones present in their oscillation power spectra. The spectra of such stars can be characterized via two global parameters, the large frequency separation ($\Delta \nu$) and the frequency at maximum power ($\nu_{\mathrm{max}}$). The former describes the regular frequency interval separating overtone modes of a given angular degree. The latter refers to the central frequency of the Gaussian-like envelope describing the visible power excess caused by the modes. These two parameters are the most readily available in the spectrum of a solar-like oscillator and, when combined with an independent measure of effective surface temperature ($T_{\mathrm{eff}}$), can be exploited to determine the mass and radius of a star to within a few percent \citep{2012ApJ...757...99S, 2012ApJ...760...32H, 2016MNRAS.460.4277G, 2016ApJ...832..121G, 2016MNRAS.462.1577Y, 2021MNRAS.501.3162L, 2022A&A...657A..31M}.

With the only requirement for the excitation of modes being the presence of an outer convection zone, solar-like oscillations have been observed in stars on the main sequence \citep[e.g.][]{2014ApJS..210....1C}, in the subgiant phase \citep[e.g.][]{2012A&A...543A..54A, 2022A&A...657A..31M}, and on the red giant branch \citep[e.g.][]{2010ApJ...713L.176B, Yu_2018, 2021MNRAS.503.4529C}. As a given star evolves through these phases, structural changes will affect the properties of the oscillations. The least evolved stars oscillate at a few thousand $\mu$Hz. This decreases as the star evolves off the main-sequence, dropping to below $\sim$100 $\mu$Hz on the red giant branch. Despite the large coverage of the Hertzprung-Russell (HR) diagram, current catalogues are  disjoint in evolutionary state. Detections are dominated by a large number of red giants and a much smaller set of main-sequence stars, with the subgiant phase only sparsely sampled. Although a decrease in numbers is expected during this phase, given their rapid evolution, observational constraints have magnified the discrepancy. 
 
Relying predominantly on space-based photometry means observations of solar-like oscillators are mostly limited to data collected by a handful of missions. Of these, the \textit{Kepler} \citep{2010Sci...327..977B} mission provides the longest time series for a large number of available targets. Observing the same patch of sky for four years, the mission monitored approximately 196,000 targets \citep[e.g.][]{2014ApJS..211....2H}. Data were collected in two modes - long and short-cadence, with the associated sampling rates corresponding to Nyquist frequency limits of 283 $\mu$Hz and 8496 $\mu$Hz, respectively. The short-cadence data spans the full range in frequency where solar-like oscillations are located. However, due to telemetry constraints, the number of targets observed in the longer cadence greatly outnumbered those in short. Indeed, of the total observed targets only a few thousand were selected for short-cadence \citep{2016ksci.rept....9T}. Therefore, of the tens of thousands of solar-like oscillators detected using \textit{Kepler} data, the vast majority are more evolved stars \citep{2013ApJ...765L..41S, 2011A&A...525A.131H, Yu_2018}. 

Although sparse in comparison, detections of solar-like oscillators were made in the short-cadence data \citep{2011Sci...332..213C}. Due to the reduced capacity for such observations, these slots were reserved for targets most suited to the main aims of the mission: the detection of exoplanets via the transit method. This led to the preferential selection of cool main sequence stars \citep{2010ApJ...713L.109B}. The combination of the selection criteria for the short-cadence observations and the Nyquist frequency in the long cadence data resulted in the asteroseismic yield lacking a significant number of subgiant stars. The largest list of such stars was constructed by \citet{2020MNRAS.495.2363L} and numbers only 50 subgiants.

 The TESS mission \citep{2014SPIE.9143E..20R} launched in 2018 and has been surveying the majority of the sky, providing an extensive database of potential solar-like oscillators. The nominal mission lasted two years and observations continue during the first extended mission, which will finish observing in September of 2022. To maximise the sky coverage, observations are made in sectors with an average length of 27.4 days. Most targets are captured in one or two sectors, while a small number of stars are located where the sectors overlap at the ecliptic poles (known as the \textit{continuous viewing zones}). Similarly to \textit{Kepler}, the majority of the stars monitored by TESS in the nominal mission were observed at 30-minute cadence (referred to as Full Frame Images, or FFIs), corresponding to a Nyquist frequency of 278 $\mu$Hz. Currently the largest systematic searches for solar-like oscillators have been performed with observations at this cadence, and therefore restricted to the more evolved stars \citep{2020ApJ...889L..34S, hon2021quick, 2021MNRAS.502.1947M, 2022MNRAS.512.1677S}. Shorter cadences are available for a smaller set of targets, with the nominal mission including a 120-second integration time (double the \textit{Kepler} short-cadence). The extended mission introduced 20-second data for a reduced target list while the FFI cadence was shortened to 10-minutes. With Nyquist frequencies of 4167 $\mu$Hz and 25000 $\mu$Hz respectively, the 120-second and 20-second cadence data allow us to detect solar-like oscillations in less evolved solar-like oscillators. 
 
 To that end, we used 120-second and 20-second TESS data to search for oscillations in stars observed during Sectors 1 to 46. Starting with a smaller set of targets that were identified as the most likely to oscillate above the 30-minute FFI Nyquist frequency \citep{2019ApJS..241...12S}, we have identified 400 candidate solar-like oscillators by eye. These were used to optimise a detection algorithm presented in \citet{Nielsen2021} (henceforth referred to as N22). We then passed the remaining stars observed during the aforementioned sectors to this tuned pipeline. Although the main aim of this work is to construct a list of solar-like oscillators, we found we could exploit the probability distributions calculated by the algorithm to measure the global properties ($\nu_{\mathrm{max}}$ and $\Delta \nu$). Therefore, we provide these values for the majority of the detected solar-like oscillators. 


\section{Target selection}\label{targets}

The full list of targets observed in 120-second cadence by TESS exceeds 300,000. The Asteroseismic Target List \citep[ATL,][]{2019ApJS..241...12S} gives some indication of which stars are most likely to be solar-like oscillators prior to running the algorithm. By separating this sample from the full set of 120-second cadence targets we can loosen detection constraints whilst keeping the required manual validation to manageable levels. To distinguish the stars in the ATL from the remaining targets observed in 120-second cadence we will refer to the latter sample as `the Large Sample'.

The ATL was constructed prior to the launch of TESS to provide a prioritised list of targets most likely to yield detections of solar-like oscillations
(\citealp{2019ApJS..241...12S}; see also \citealp{2021PASP..133i5002F, 2021ApJ...915...19G}). Aimed at 120-second cadence data, the list was restricted to stars which would oscillate above the 30-minute FFI Nyquist frequency. To select targets the authors employed asteroseismic scaling relations for $\nu_{\mathrm{max}}$ \citep{2016ApJ...830..138C}. This allowed them to locate stars in the TESS field-of-view predicted to have $\nu_{\mathrm{max}}$ > 240 $\mu$Hz. Calculating the expected power excess caused by the modes, the authors estimated the probability that the oscillations would be detectable.
Only those targets with at least a 5\% probability of making a detection were retained, which constituted $\approx$ 25,000 targets. Of these, 11,220 had been observed at the time of this work. In the following analysis, values of parallax and $T_{\mathrm{eff}}$ (required by the detection algorithm) were taken from the ATL. The ATL used parallaxes from Gaia data release 2 \citep[GDR2,][]{refId0}, supplemented at bright magnitudes with values from the eXtended Hipparcos Catalogue \citep[XHIP,][]{2012}. Effective temperatures in the ATL were computed from a polynomial in dereddened (B-V) colour, using coefficients according to \citet{2010AJ....140.1158T}. 

The Large Sample consists of the remaining 120-second cadence targets. We selected stars brighter than 11th magnitude in 2MASS $\Ks$ magnitude, and used $T_{\mathrm{eff}}$ from the TESS Input Catalogue \citep[TIC, ][]{2019AJ....158..138S} to restrict to the range 4500K < $\Teff$ < 6500K. This includes the typical ranges in $T_{\mathrm{eff}}$ of stars from the main sequence to the red giant branch, and removes stars that are likely too faint for modes to be visible \citep{2017ApJ...835...83S}. In this set we analysed light curves for 255,089 stars. We applied the same cuts to all of the targets observed in 20-second cadence, which yielded light curves for 6157 stars. Parallaxes for both sets were again drawn from GDR2.

For reference, we also identified targets with a published detection of solar-like oscillations and some measure of the global asteroseismic parameters.  A set of 13 such stars was produced, which is shown in Table $\ref{tab:Lit Osc}$ and referred to in the following as the `literature sample'. We prioritized targets oscillating at frequencies above the \textit{Kepler} long-cadence Nyquist as per the main aim of the catalogue. Parallaxes and effective temperatures were drawn from GDR2. 

\begin{table*}
	\centering
	\begin{threeparttable}
	\caption{Global asteroseismic parameters of stars in the literature sample.}
	\label{tab:Lit Osc}
	\begin{tabular}{lcccr} 
		\hline
		Name & TIC & $\nu_{\mathrm{max}}$ ($\mu$Hz) & $\Delta \nu$ ($\mu$Hz) & Source \\
		\hline
		HD 19916 & 200723869 & 1188 $\pm$ 40  & 61.4 $\pm$ 1.5 & TESS 120-second \tnote{1}\\
		HD 222416 & 441462736 & 430 $\pm$ 18 & 28.94 $\pm$ 0.15 & TESS 120-second \tnote{2}\\
		$\lambda^{2}$ For & 122555698 & $\approx$ 1280  & 69.76 $\pm$ 0.23 & TESS 120-second \tnote{3}\\
		HD 212771 & 12723961 & 226.6 $\pm$ 9.4 & 16.25 $\pm$ 0.19  & TESS 30-minute FFI \tnote{4}\\
		HD 222076 & 325178933 & 203.0 $\pm$ 3.6 & 15.60 $\pm$ 0.13 & TESS 120-second \tnote{5}\\
		94 Aqr & 214664574 & 875 $\pm$ 12 & 50.2 $\pm$ 0.4 &TESS 120-second \tnote{6}\\
		$\gamma$ Pav & 265488188 & 2693 $\pm$ 95 & 119.9 $\pm$ 1.0 & TESS 20-second \tnote{7}\\
		$\pi$ Men & 261136679 & 2599 $\pm $69 & 116.7 $\pm$ 1.1 & TESS 20-second \tnote{7}\\
		$\nu$ Ind & 317019578 & & 25.08 ± 0.10 & TESS 120-second  \tnote{8}\\
		$\beta$ Hyi & 267211065 & $\approx$ 1000 & 57.24 $\pm$ 0.16 & HARPS and UCLES, WIRE \tnote{9}\\
		$\mu$ Ara & 362661163 & $\approx$ 2000 & 89.68 $\pm$ 0.19 & HARPS \tnote{10}\\
		$\mu$ Her & 460067868 & 1216 $\pm$ 11 & 64.2 $\pm$ 0.2 & SONG \tnote{11} \\
		$\alpha$ Men & 141810080 & 3134 $\pm$ 440 & 140 $\pm$ 2 & TESS 20-second \tnote{12}\\
		\hline
	\end{tabular}
	\begin{tablenotes}
	References.
    \item[1] \citet{2021MNRAS.502.3704A}
    \item[2] \citet{2019AJ....157..245H}
    \item[3] \citet{2020A&A...641A..25N}
    \item[4] \citet{2019ApJ...885...31C}
    \item[5] \citet{2020ApJ...896...65J}
    \item[6] \citet{2020ApJ...900..154M}
    \item[7] \citet{2022AJ....163...79H}
    \item[8] \citet{Chaplin_2020}
    \item[9] \citet{2007ApJ...663.1315B}, \citet{2007CoAst.150..147K}
    \item[10] \citet{2005A&A...440..609B}
    \item[11] \citet{2017ApJ...836..142G}
    \item[12] \citet{2021ApJ...922..229C}
    \end{tablenotes}
    \end{threeparttable}
\end{table*}

\section{Data selection}\label{data}

We used detrended light curves produced by the TESS Science Processing Operations Center (SPOC) pipeline \citep{2016SPIE.9913E..3EJ}, which carries out the simple aperture photometry and removes instrumental trends. We used light curves recorded in 120-second cadence, except when 20-second data were available. In the literature sample, we used 120-second cadence light curves for all but three stars ($\gamma$ Pav, $\pi$ Men and $\alpha$ Men), where we used 20-second cadence light curves. We used the open source package \texttt{Lightkurve} \citep{2018ascl.soft12013L} to stitch sectors together, and remove flux values exceeding 5$\sigma$. 

As TESS observes in 27.4 day sectors, there are gaps present in the light curves. During the nominal mission the northern and southern hemispheres were observed for 13 sectors respectively, amounting to a year in each. The extended mission returned to the southern hemisphere, meaning the light curves of some targets contain year-long gaps. Both leaving the gaps and methods to fill the gaps (such as linear interpolation, \citet{2015ApJ...809L...3S}) introduce strong correlations between frequency bins. Assuming mode lifetimes follow the relation given by \citet{2012A&A...537A.134A} \citep[see also][]{2017ApJ...835..172L}, at $\Teff$ = 5000K, we expect mode lifetimes on the order of weeks. Therefore, if a star is observed in the nominal mission and then a year later in the extended, we expect the modes have been re-excited so that the variability in the time series is no longer correlated. Hence we removed the gap in the data by shifting the timestamps. As in N22, gaps larger than 50 days were treated this way. Gap closing in this manner is not the optimal approach and does alter the line profiles, which would impact measurements of individual frequencies, as was discussed in \citet{Bedding_2022}. However, as noted in N22, the inclusion of gaps significantly increased the false positive rate (see N22 Fig. A.1) necessitating closing the gaps for our detection tests. Furthermore we report only the global seismic parameters, and do not measure individual frequencies. 

We then used the \texttt{Lightkurve} package to produce a power density spectrum via the `fast' Lomb-Scargle method \citep{1976Ap&SS..39..447L, 1982ApJ...263..835S, 1989ApJ...338..277P}.

\section{Detection of solar-like oscillators}\label{detection}

In the following section we briefly review the methods used in the detection algorithm of N22, before discussing the detections made in each set of targets. The detection test consists of two modules, which exploit different properties of solar-like oscillators:
 \begin{enumerate}
     \item \textbf{Power excess test:} The first module uses the power spectral density of the time series. Given the assumption that the noise in each frequency bin follows a $\chi^2$ distribution with 2 degrees of freedom, the probability that only noise is present in a given bin is calculated (the H$_{0}$ probability). The probability that an envelope is present is then computed via a prediction of the expected power in a hypothetical envelope centered on each frequency bin (the H$_1$ probability). The prediction is calculated via the methods of \citet{2011ApJ...732...54C} and \citet{2019ApJS..241...12S}, requiring $\Teff$. Prior information on $\nu_{\mathrm{max}}$ is used, guided by parallax and 2MASS $\Ks$-band magnitude.
     \item \textbf{Large separation test:} The second module looks for the signature of the regularly spaced overtones. This is achieved using the methods described in \citet{2009A&A...508..877M}, who used the autocorrelation function (ACF) of the time series. A band-pass filter is placed on the power spectrum at a test frequency and the ACF of this filtered time series is calculated via an inverse Fourier transform. Repeating this for other test frequencies, we produce a 2D ACF in test frequency (a proxy for $\nu_{\mathrm{max}}$) and lag (related to the large frequency separation via $\tau=1/\Delta \nu$). This 2-dimensional map is collapsed along the lag axis to produce a 1-dimensional distribution in frequency. To evaluate whether the probability the observed collapsed ACF is inconsistent with noise, N22 approximate the response due to noise by a $\Gamma$ distribution.
 \end{enumerate}
 
Using 400 solar-like oscillators and non-detections manually identified from the ATL, N22 established the performance of the pipeline for a range of detection thresholds. They found the pipeline was able to attain a true positive rate of  94.7\% and a false positive rate of 8.2\% when asking for a response at least one of the two modules\footnote{This is achieved when taking a threshold of 0.77 on the power excess module and 0.73 on the frequency spacing module.}. 

In the following sections we discuss the detections made in the four samples via this algorithm (see Table \ref{tab:big table}). All of the reported solar-like oscillators have been manually vetted to check for false positives. We retain only targets in which we are confident we can identify the presence of oscillations, prioritizing a reduced false positive rate over maximising the yield. This may cause an under-representation of targets with very low signal-to-noise. A break down of the total counts, and which sample they belong to, can be seen in Table \ref{tab:Det}. Cross referencing with NASA's Exoplanet Archive \footnote{\url{https://exoplanetarchive.ipac.caltech.edu}}, we found 28 of the stars in our catalogue are confirmed planet hosts. The majority of these stars have not yet been studied asteroseismically. Asteroseismic inferences on these targets are reserved for an upcoming work. We also cross-referenced with the \textit{Ninth Catalogue of Spectroscopic Binary Orbits} \citet[SB9,][]{2004A&A...424..727P} discovering a further 25 stars are components in spectroscopic binary systems. 

\begin{figure}
    \centering
    \includegraphics[width=1\linewidth]{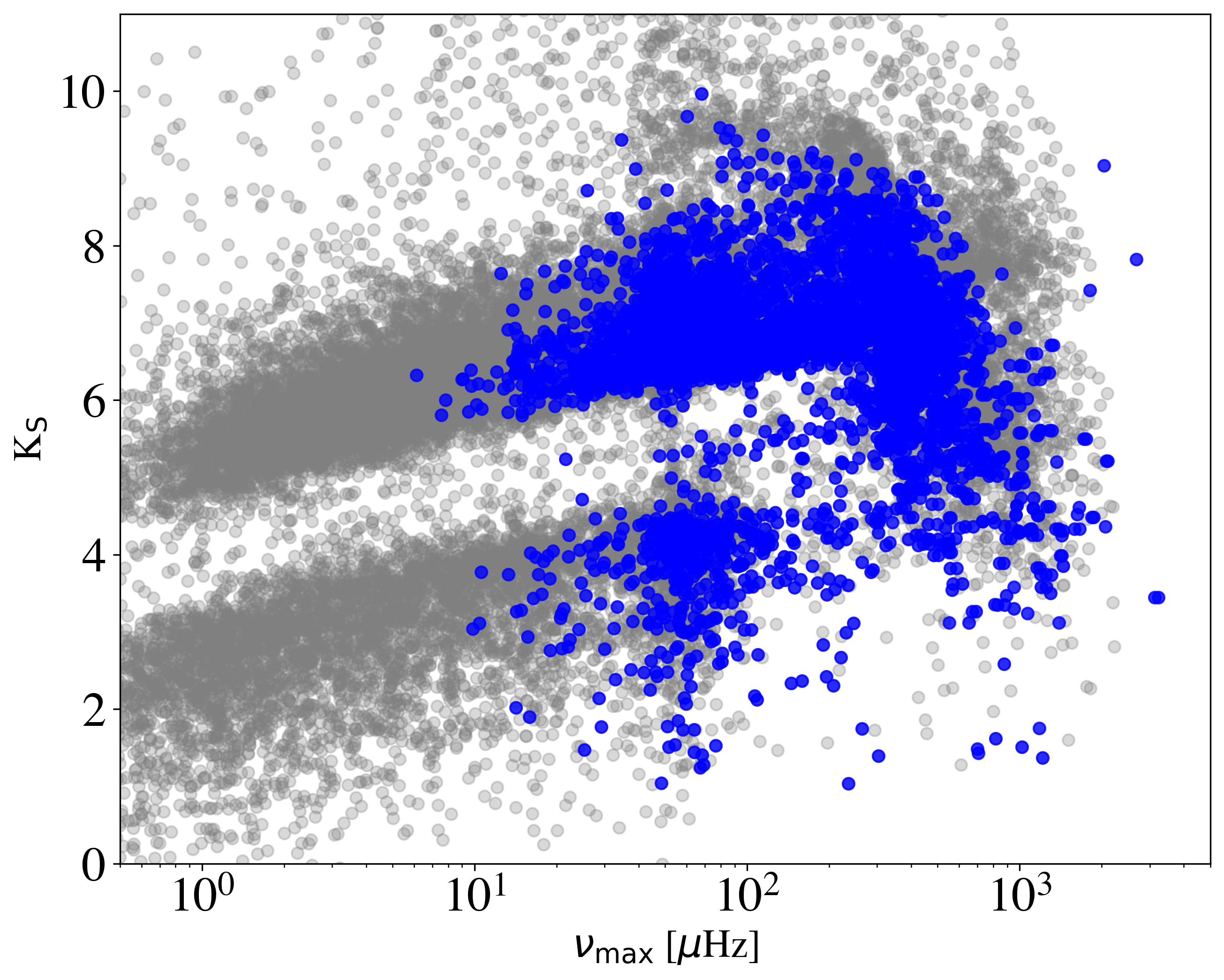}
    \caption{2MASS $K_{\mathrm{S}}$ Magnitude and predicted $\nu_{\mathrm{max}}$ for stars observed in 120-second cadence with a detection probability exceeding 10\% (grey). The targets which we identified as solar-like oscillators are marked in blue.}
    \label{fig:mag limits}
\end{figure}

Fig. \ref{fig:mag limits} shows the value of 2MASS $K_{\mathrm{S}}$ magnitude against the predicted value of $\nu_{\mathrm{max}}$. In grey we show targets observed in 120-second cadence with a detection probability greater than 10\% using the methods described by \citet{2011ApJ...732...54C} and \citet{2019ApJS..241...12S}. The gap in the grey population starting at K$_{\mathrm{S}} \approx 4$ is present in the full sample of short cadence targets and not enforced by the probability cut, which only puts upper limits on $\nu_{\mathrm{max}}$ and magnitude. It is likely the result of the TESS target selection process, which consists of selecting stars from a number of lists including cool dwarfs, known planet hosts, bright stars, hot subdwarfs and guest investigator targets \citep{2018AJ....156..102S, 2019AJ....158..138S}. We did not detect solar-like oscillations in any targets with $\nu_{\mathrm{max}}$ < 5 $\mu$Hz, regardless of magnitude. As discussed in N22, the predicted mode amplitude used in the power excess module included the observed decrease near the red edge of the $\delta$ Scuti instability strip. This was done via a factor which depends on $\nu_{\mathrm{max}}$ and $T_{\mathrm{eff}}$, so that at a given temperature amplitudes decrease as a function of frequency \citep{2011ApJ...732...54C}. The correction was calibrated using main-sequence stars and therefore may not be appropriate for the most evolved targets. Indeed, at $T_{\mathrm{eff}} = 4800$K the factor decreases to approximately zero at frequencies below 5 $\mu$Hz, which suppresses detections in the power excess module.  In addition, from the approximate relation between $\nu_{\mathrm{max}}$ and $\Delta \nu$ \citep[][see Eq. \ref{eqn: scale}]{2009MNRAS.400L..80S} at a $\nu_{\mathrm{max}}$ of 5 $\mu$Hz we would expect $\Delta \nu$ to be below 1 $\mu$Hz. This is approaching the resolution in a single sector of TESS data (0.4 $\mu$Hz). Therefore, detections in the repeating pattern module are also increasingly unlikely.

\begin{table*}
    \centering
    \caption{Seismic parameters}
    \csvautobooktabular[{table head = \hline TICID & No. Sectors &  RP & PE & $\nu_{\mathrm{max}}$ ($\mu$Hz) &  $\sigma(\nu_{\mathrm{max}})$ ($\mu$Hz) & $\Delta \nu$ ($\mu$Hz) & $\sigma(\Delta \nu)$ ($\mu$Hz) &  sample &  Flag\\\hline\label{tab:big table}}]{example_of_submission.csv}
    \tablefoot{Catalogue of seismic parameters for detected solar-like oscillators. The full table is available in online materials. Quantities RP and PE track which modules the target produced a detection in. Flag is `SB9' for targets in the \textit{Ninth Catalogue of Spectroscopic Binary Orbits} \citep{2004A&A...424..727P} and `PH' for targets which are confirmed planet hosts according to NASA's Exoplanet Archive.}
\end{table*}

\subsection{Literature Sample}
Of the 13 solar-like oscillators drawn from the literature, the algorithm flagged a detection in both modules for 11 stars. $\mu$ Ara produced a response in the power excess module, but not the repeating pattern. Oscillations in this star have thus far only been detected in Doppler velocity \citep{2005A&A...440..609B}. As the signal from granulation is lower relative to the modes in velocity measurements than in photometry \citep{BasuSarbani2017Ada:}, the single module response is likely just an effect of the decreased signal-to-noise. The remaining star, HD19916, did not produce a flag in either module.  Although oscillations in HD19916 have been detected in TESS 120-second cadence data the authors note that a custom aperture had to be used, expanding to include more of the stellar flux \citep{2021MNRAS.502.3704A}. To maintain consistency with the rest of our catalogue, we did not mimic this approach. We note that the stars with detections reported in 20-second data ($\gamma$ Pav, $\pi$ Men and $\alpha$ Men) produced flags in both modules. However, if we use the available 120-second data for the same stars, one is not detected at all ($\pi$ Men) and the others are only detected in the power excess module, despite oscillating at frequencies well below the corresponding Nyquist frequency limit. The improvement made by the 20-second data was highlighted by \citet{2022AJ....163...79H}.

\subsection{Asteroseismic Target List}\label{ATL detect}

To construct the set of stars used to establish the performance of the detection algorithm in N22, a manual inspection of the 11,220 spectra discussed in Sect. \ref{targets} was performed. On the construction of the the sets of 400 oscillators and 400 `non-oscillators', several stars fell into the category of targets where, although some excess power was present in the spectrum, we were unable to unambiguously classify the target as a solar-like oscillator. Since that work was done, new sectors of data had become available, which could facilitate unambiguous classification. We therefore re-ran the algorithm on the full set of 11,220 spectra. With the testing set of oscillators from N22 removed, 2651 stars were flagged in just one module and 490 in two. Of the single module detections, the vast majority were false positives presenting some large non-solar type signal rather than solar-like oscillations (such as periodic dips caused by a transit, eclipse, or classical oscillations). Such stars were not included in the metrics stated in N22. Including these stars in the false-positive metric for a single module response increases the percentage to $\approx$20\%. In total (with the testing set included) we detected 494 solar-like oscillators with responses in both modules, and another 258 with a single module response.

\subsection{Large Sample}

For the 255,089 stars for which we analysed light curves, we expect a false positive rate of 8.2\%. Assuming the majority are not solar-like oscillators, this would equate to false positives in the range of $\approx$20,000. We found 37,250 flagged in at least one module, therefore we took the more conservative approach and performed a manual inspection of stars that produced a flag in both modules. Of the 5,781 stars which produced flags in both modules, we found 2,927 clear solar-like oscillators. We have retained the list of single-module responses, but reserve releasing it until they have been manually vetted, to avoid confusion. Unlike the ATL sample, we have relaxed the requirement that $\nu_{\mathrm{max}}$ exceeds 240 $\mu$Hz. This gives a set of targets which are cooler on average, and more strongly peaked in magnitude (see Fig. \ref{fig:Kmag big test} and \ref{fig:Teff big test}).

\subsection{20-second cadence}
Of the 6157 stars in this set, the algorithm produced a single module response for 1585, and a double module response for 421. Upon visual inspection of these targets, we were able to clearly identify 490 solar-like oscillators.

\begin{table}
    \begin{center}
    \begin{threeparttable}
	\caption{Detection counts in each sample. `Double' refers to cases where the star flagged a detection in both the power excess and repeating pattern modules. `Single' refers to cases where the star flagged in one module only.}
	\label{tab:Det}
	\begin{tabular}{p{0.07\textwidth}p{0.07\textwidth}p{0.07\textwidth}p{0.07\textwidth}}
		\hline
		Sample & Double & Single & Total \\
		\hline
		Literature & 11 & 1 & 12\\
		ATL & 494 & 258 & 752\\
		Large & 2927 & - & 2927\\
		20-sec & 288 & 198& 486\\
		\hline
		Total & 3720 & 457 & 4177\\
        \hline
	\end{tabular}
	\end{threeparttable}
	\end{center}
\end{table}

\begin{figure}
    \centering
    \includegraphics[width=1\linewidth]{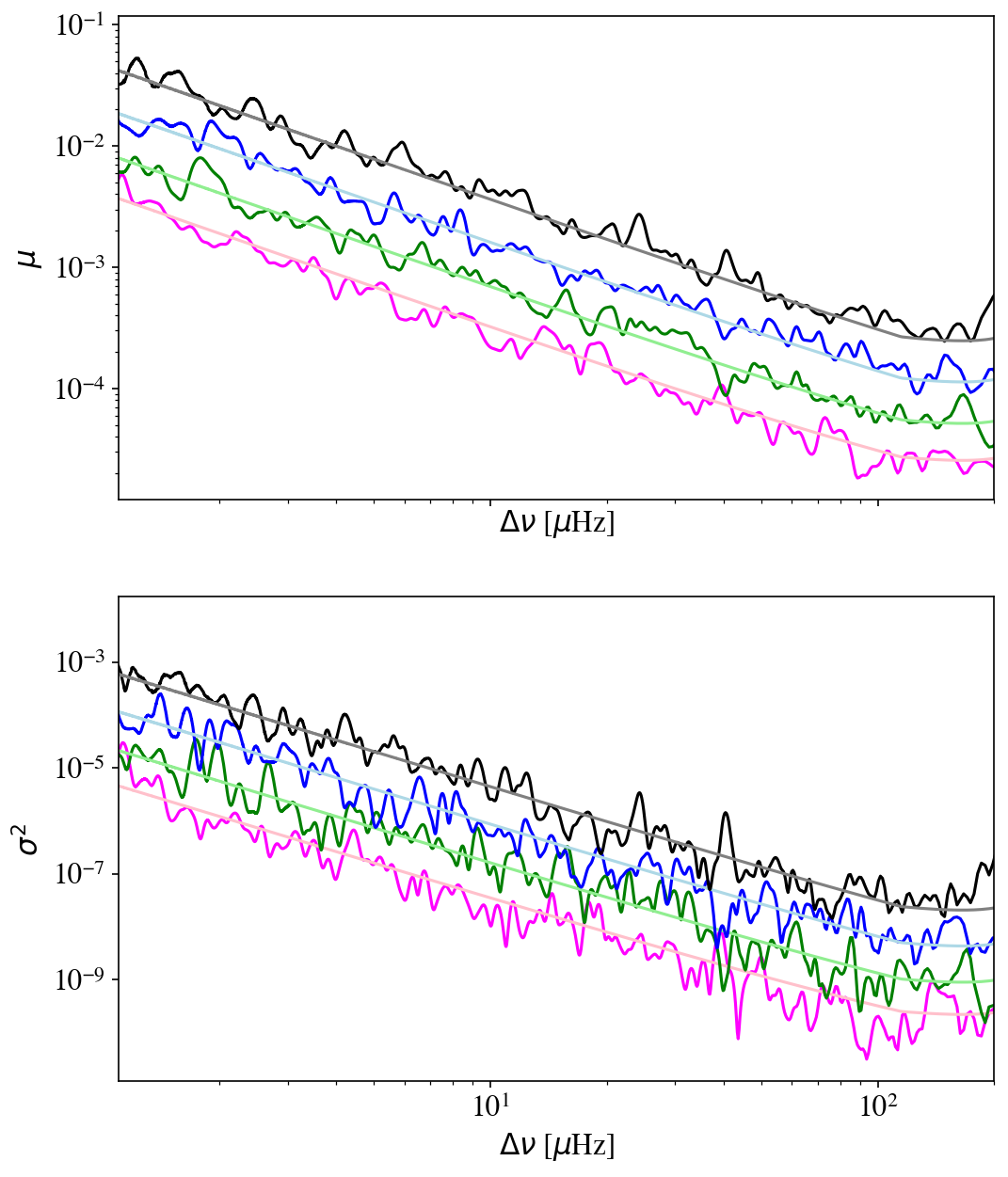}
    \caption{Top panel: Mean simulated collapsed ACF as a function of $\Delta \nu$ for filtered white noise. Colours represent time series of different lengths, with one sector in black, four in blue, nine in green and twelve in magenta. Pale lines show the predictions for each length according to Eq. \ref{eqn: model}. Bottom panel: Variance on the simulated collapsed ACF presented in the top panel.}
    \label{fig:Modelled mean}
\end{figure}

\begin{figure}
    \centering
    \includegraphics[width=0.95\linewidth]{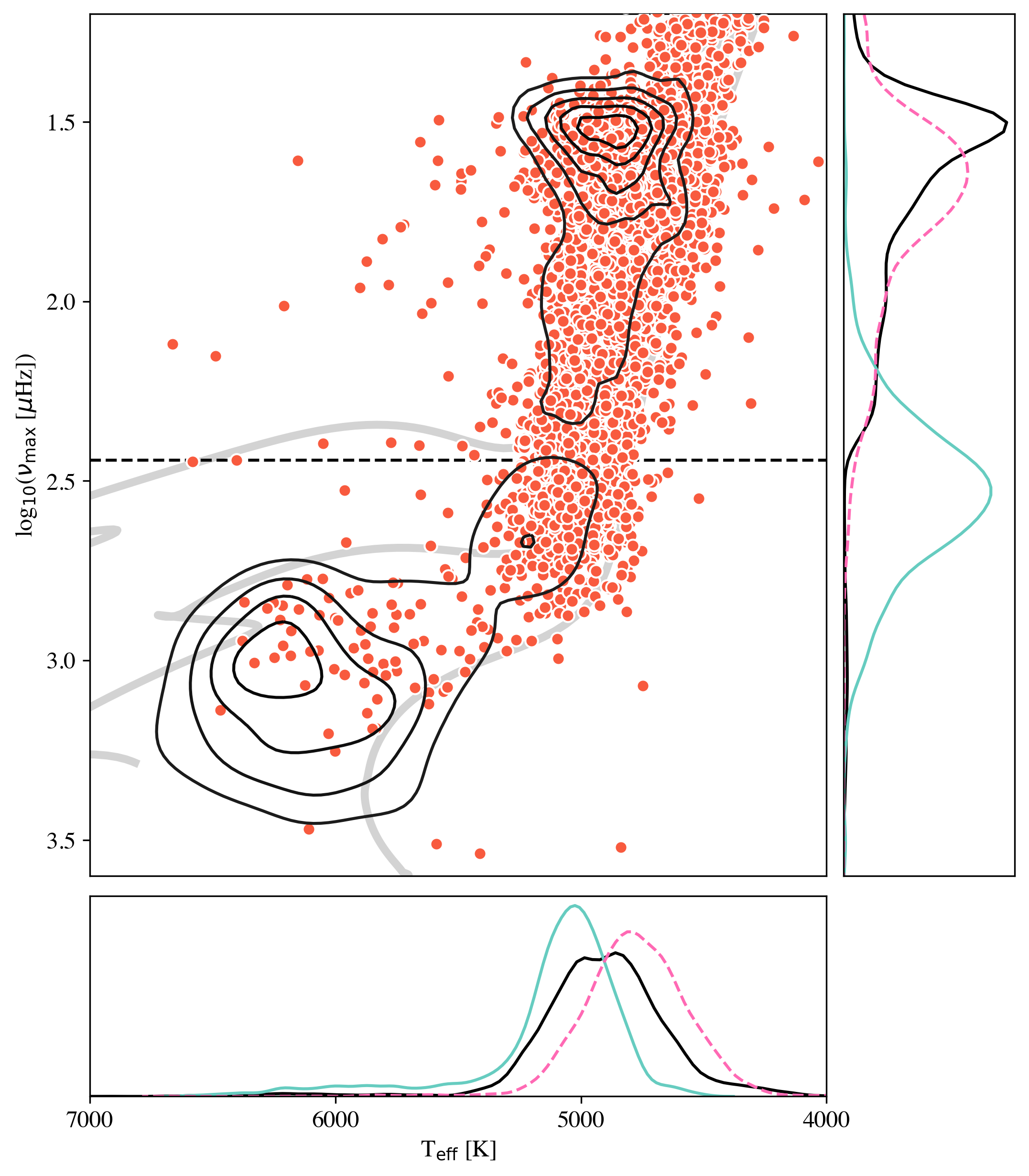}
    \caption{Main panel shows an asteroseismic HR diagram for stars in all samples (orange circles). Effective temperatures have been drawn from the TIC in all cases, to maintain consistency. Contour lines represent measurements from \textit{Kepler} data reported in \citet{Yu_2018}, \citet{2017ApJ...835..172L} and \citet{2017ApJS..233...23S}, with effective temperatures from GDR2. The horizontal dashed line represents the \textit{Kepler} long cadence Nyquist frequency. Three stellar tracks at masses 1.0, 1.5 and 2.0 M$_{\odot}$ generated by MIST \citep{2016ApJ...823..102C} are shown in grey. The distributions in $\Teff$ and $\nu_{\mathrm{max}}$ are shown in the bottom and right panels, respectively. Here, the catalogue is split into the ATL set in turquoise and the Large Sample in pink (dashed) with the \textit{Kepler} distribution shown in black.}
    \label{fig:Seis HR}
\end{figure}

\begin{figure*}
    \centering
    \includegraphics[width=0.9\linewidth]{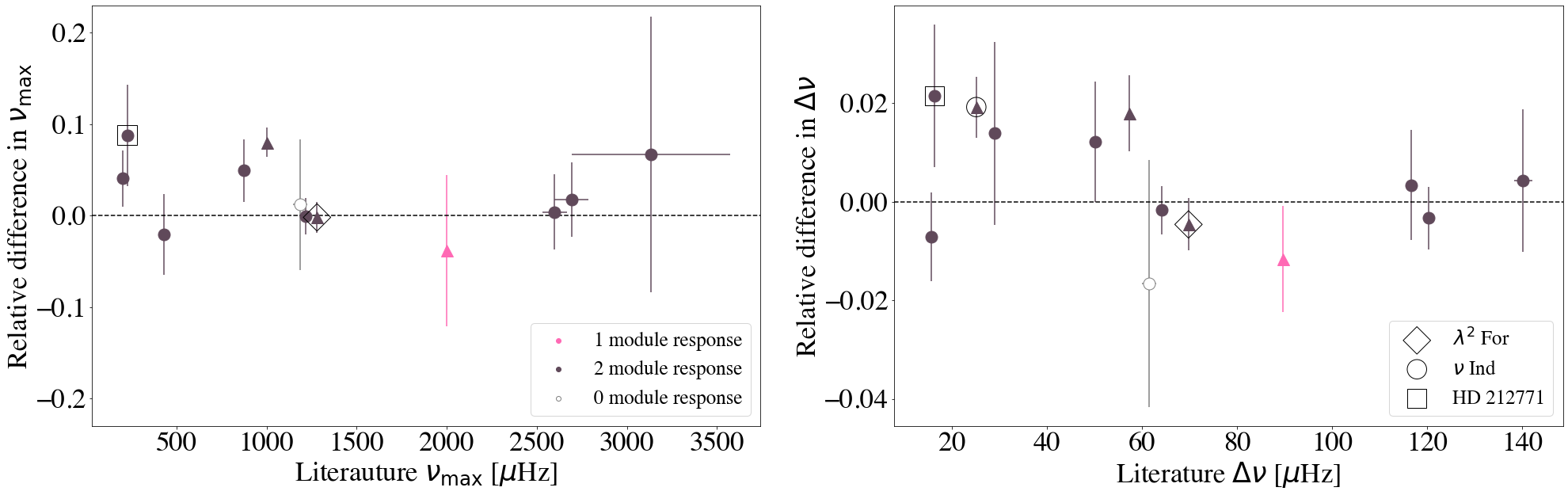}
    \caption{A comparison of the global asteroseismic parameters measured by our algorithm to those reported in the literature. Stars are coloured by the number of modules in which they produce a flag; pink for a single module, brown for two modules and an open grey marker for none. Triangles represent stars for which no uncertainty on $\nu_{\mathrm{max}}$ was reported in the literature. Targets $\lambda^{2}$ For, $\nu$ Ind and HD212771 which are discussed in the text are marked with a diamond, circle and square respectively. The left and right panels show the fractional difference between the values of $\nu_{\mathrm{max}}$ and $\Delta \nu$ as measured by the algorithm versus literature value respectively.}
    \label{fig:Residuals}
\end{figure*}

\begin{figure*}
    \centering
    \includegraphics[width=0.8\linewidth]{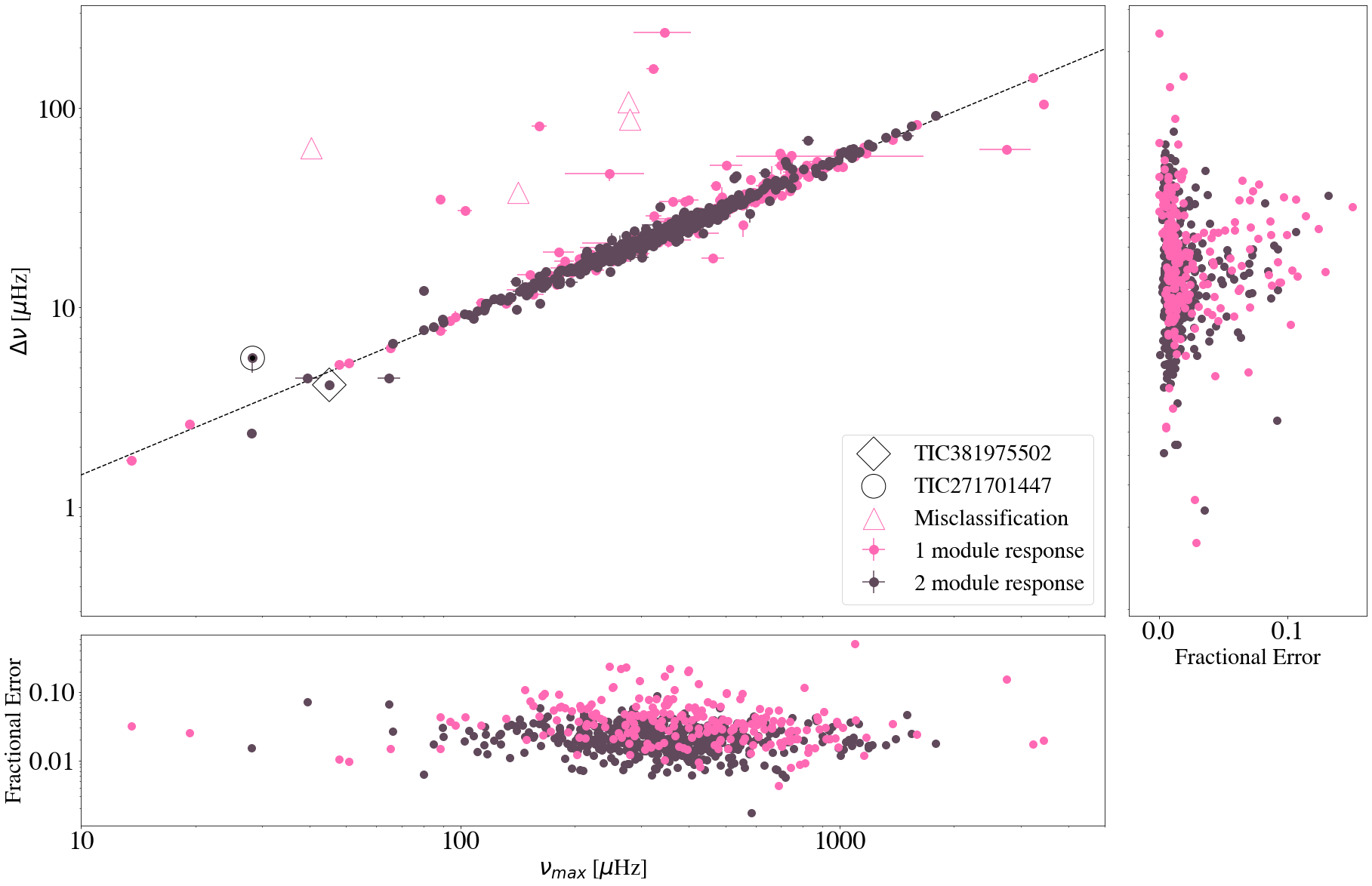}
    \caption{The main panel shows $\Delta \nu$ as a function of  $\nu_{\mathrm{max}}$ measured by the algorithm for validated solar-like oscillators from the ATL set. Stars producing flags in only one module are shown in pink, those producing flags in both modules are shown in brown. Triangles mark stars that were likely misclassified as solar-like oscillators. Targets TIC381975502 and TIC271701447 which are discussed in the text have been marked with a diamond and circle. The black dotted line shows Eq. \ref{eqn: scale}. Additional panels show the fractional uncertainties on $\Delta \nu$ and $\nu_{\mathrm{max}}$.}
    \label{fig:dnu numax 733}.
\end{figure*}

\section{Global asteroseismic parameters}\label{globals}
\begin{table}
    \begin{center}
	\caption{Parameters for Eq. \ref{eqn: model} and \ref{eqn:var model}.}
	\label{tab:fits}
	\begin{tabular}{lcr}
		\hline
		Parameter & Value\\
		\hline
		$B$ & 1.54\\
		$\alpha$ & 0.65\\
		$\beta$ & 0.36 \\
        $c$ & 0.34 \\
		\hline
	\end{tabular}
	\end{center}
\end{table}

Alongside enabling detection, the probability distributions calculated in the detection process allow us to measure the global asteroseismic parameters $\Delta \nu$ and $\nu_{\mathrm{max}}$. There are already a number of pipelines dedicated to measuring these parameters via different methods \citep{2009CoAst.160...74H, 2010MNRAS.402.2049H, 2010A&A...522A...1K, 2010A&A...511A..46M, 2014A&A...568A..10G, 2017MNRAS.466.3344E, 2019ApJ...884..107Z}. However, as the main aim of this work is the construction of a list of solar-like oscillators, a full comparison between our method and such alternatives is reserved for future work. 

The probability distribution as a function of frequency calculated by the frequency spacing method (see Sect. \ref{detection}) was normalized to unit integral over the spectrum, producing a probability density. The 50th percentile of this was used to measure $\nu_{\mathrm{max}}$, with the 16th and 84th percentiles giving the confidence interval. 

N22 only required the ACF collapsed along the lag ($\tau$) axis to perform a detection. To determine $\Delta \nu$ we instead collapsed along test frequency. Rather than summing the ACF for all test frequencies at a given lag, we exploited the approximate relation between $\nu_{\mathrm{max}}$ and $\Delta \nu$ \citep{2009MNRAS.400L..80S},
 
 \begin{equation}\label{eqn: scale}
     \frac{\Tilde{\nu}_{\mathrm{max}}}{\nu_{\mathrm{max, \odot}}} =  \bigg(\frac{\Delta \nu}{\Delta \nu_{\odot}}\bigg)^{a} ,
 \end{equation}
 where we took the value $a$ = 0.791 as in N22. This estimate of $\nu_{\mathrm{max}}$ at a given $\Delta \nu$ ($\Tilde{\nu}_{\mathrm{max}}$) allowed us to restrict the range of frequencies summed. Accordingly, we only summed bins in the range $|\nu_{\mathrm{max}} - \Tilde{\nu}_{\mathrm{max}}| < 0.2 \Tilde{\nu}_{\mathrm{max}}$. 
 
 To calculate the expectation from a spectrum devoid of oscillations we utilized 10$^{3}$ white noise realizations. Similarly to N22 we found that, for filtered white noise, the noise statistics can be well approximated by a $\Gamma$ distribution in lag. The mean of this distribution can be described by the empirical relation
\begin{equation}\label{eqn: model}
    \mu(\tau) = A(B + \tau^{\alpha}/N_{\nu}^{\beta}),
\end{equation}
where $N_{\nu}$ is the number of frequency bins included in the calculation of the ACF at a given $\tau$. Utilizing the \texttt{emcee} package \citep{2013PASP..125..306F} we fitted for the parameters $B$, $\alpha$ and $\beta$, the results of which can be found in Table \ref{tab:fits}. The parameter $A$ is a calibration constant which depends on the time series length and was determined on a star-by-star basis. Using Eq. \ref{eqn: scale} to estimate the value of $\Delta \nu$ ($\widetilde{\Delta\nu}$) given the measured value of $\nu_{\mathrm{max}}$, we masked the ACF in the range 0.7$\widetilde{\Delta\nu}$ < $\Delta \nu$ < 1.3$\widetilde{\Delta\nu}$. The calibration factor $A$ was then estimated by the ratio of the modelled to observed ACF in the first 5 $\mu$Hz and final 50 $\mu$Hz (with the latter range accounting for the decrease in frequency resolution at small lag). 

We found that the variance in the collapsed ACF can be approximated by,
\begin{equation}\label{eqn:var model}
    \sigma(\tau)^{2} = c\mu(\tau)^{2},
\end{equation}
with the value of $c$ determined by a fit to the white noise simulations  (see Table \ref{tab:fits}). A comparison of the predictions from Eq. \ref{eqn: model} and \ref{eqn:var model} to simulations of different time series length can be seen in Fig. \ref{fig:Modelled mean}. We also tested the model on data binned to different effective lengths. The net effect of the binning is an additional multiplicative factor, which is accounted for in the calibration.

We used Eq. \ref{eqn: model} and \ref{eqn:var model} to establish the probability ($P_{\Delta \nu}$) that the collapsed ACF ($r$) at a given value of $\tau$ is inconsistent with noise. Logarithmic probabilities were used for numerical stability. Given that the envelope will cause an excess above the mean, we can label any divergences below the mean as noise. Therefore the natural choice is the survival function,

\begin{equation}
    \log P_{\Delta\nu} = -\log\left(\int_{r}^{\infty}\frac{\beta^{\alpha}}{\gamma(\alpha)}r^{\prime,\alpha - 1}\mathrm{exp}(-\beta r^{\prime})dr^{\prime}\right),
\end{equation}
where the shape parameter is $\alpha(\tau) = (\mu(\tau)/\sigma(\tau))^{2}$ and the scale parameter $\beta(\tau) = \mu(\tau)/\sigma(\tau)^{2}$. Normalizing $P_{\Delta \nu}$ to unit integral over the $\Delta \nu$ axis produces a probability density. The 50th percentile of this was used to measure $\Delta \nu$, with the 16th and 84th percentiles giving the confidence interval.

In the following sections we discuss the values of $\Delta \nu$ and $\nu_{\mathrm{max}}$ in each of our samples. We utilize our literature sample to briefly comment on the robustness of our methods in TESS data compared to results produced largely from bespoke analysis of individual stars. We then go on to discuss the results in the remaining new detections. A summary of the catalogue can be seen in Fig. \ref{fig:Seis HR}. We note that in some targets, values of $\Delta \nu$ may be reported despite the star not producing a flag in the repeating pattern module. To report a detection we require that the repeating pattern merit function exceeds a threshold that was chosen as the best balance between the false positives and false negatives. Therefore, a star could produce a measurable response in the collapsed ACF, while the merit function peaks just below the selected threshold. Accordingly, we only removed measurements from the final catalogue that were manually identified as clear outliers in the $\nu_{\mathrm{max}}$-$\Delta \nu$ plane. The values that have been removed shown in Fig \ref{fig:bad globals} of the appendix.

\subsection{Literature Sample}

Of the 13 targets making up the sample, 9 have published $\nu_{\mathrm{max}}$ values with uncertainties. In the remaining stars, the authors focussed on determining individual frequencies rather than global parameters, and so estimates of $\nu_{\mathrm{max}}$ without uncertainties were published. A comparison of the literature values to those measured by our method can be seen in Fig. \ref{fig:Residuals}. On average, the measured values of $\nu_{\mathrm{max}}$ are larger than those reported in the literature by $\approx$ $2.5\%$. The star with largest fractional difference is HD 212771, where our value of $\nu_{\mathrm{max}}$ is larger by $\approx$ $9\%$. The literature value was measured using FFI data processed by the TESS Asteroseismic Science Operations Center (TASOC) pipeline \citep{2021AJ....162..170H}. From visual inspection of the signal-to-noise spectrum of HD 212717 we found the envelope extended beyond the FFI Nyquist frequency. The attenuation caused by the sampling integration causes a decrease in power near the Nyquist frequency, which could have caused an underestimate on $\nu_{\mathrm{max}}$ in \citet{2020ApJ...889L..34S}. Visual inspection of the power spectra confirms our higher value is more accurate. 

In total 11 stars have measurements of $\Delta \nu$ in the literature. In the case of $\nu$ Ind, the most recent asteroseismic study was made in \cite{Chaplin_2020}, where the authors used a single sector of TESS data to fit individual modes. We measured $\Delta \nu$ from the gradient of a linear fit to the radial mode frequencies as a function of order \citep{2011ApJ...743..161W}. We took the same approach for $\lambda^2$ For where the authors also did not provide estimates of $\Delta \nu$ \citep{2020A&A...641A..25N}. A comparison of the values with those from the algorithm can be seen in the right panel of Fig. \ref{fig:Residuals}. The agreement is better than for $\nu_{\mathrm{max}}$, with a mean fractional difference on $\Delta \nu$ of $0.24\%$. 

\subsection{Asteroseismic Target List}
 
Of the 752 validated solar-like oscillators, we report both $\Delta \nu$ and $\nu_{\mathrm{max}}$ for 739. In the majority of cases the determination of the two goes hand-in-hand, with both coming from the autocorrelation function collapsed along the relevant axis. However it is possible to detect the envelope without a signal from the frequency spacing. 

In order to assess the quality of the measured $\nu_{\mathrm{max}}$ and $\Delta \nu$ values, we exploit the approximate scaling relation between the two (Eq. \ref{eqn: scale}).  Although there is a slight mass dependence in the exponent $a$ \citep{2009MNRAS.400L..80S}, the general trend remains such that stars disagreeing significantly with the rest of the population may indicate an error in one (or both) of the measured values.

Fig. \ref{fig:dnu numax 733} shows the relation between the values of $\nu_{\mathrm{max}}$ and $\Delta \nu$ for the targets in the ATL sample. The majority of the stars with detections in both modules follow Eq. \ref{eqn: scale}. The star with smallest $\Delta \nu$ is TIC 381975502 (CD-56 1110). Here, a background eclipsing binary introduced several harmonic peaks in the power spectral density at low frequency. These appear in the ACF as a large response at a test frequency corresponding to the frequency of the orbital harmonics, resulting in the algorithm incorrectly assigning both $\Delta \nu$ and $\nu_{\mathrm{max}}$. This is also the case for TIC 271701447 (HR 4749; HD 108570). The values of $\nu_{\mathrm{max}}$ and $\Delta \nu$ for these targets have been removed from the final catalogue, while IDs have been retained.

 For targets detected in just one module there is a larger scatter about the scaling relation as shown on Fig. \ref{fig:dnu numax 733}. Using an exponent on Eq. \ref{eqn: scale} of $a=0.791$ as calculated in N22, $\Delta \nu$ for 13 targets differs by more than 30\% from the value predicted by the scaling relation and measured value of $\nu_{\mathrm{max}}$. We performed a manual inspection of these stars and found four are likely misclassifications. A further two passed the detection threshold, both at the envelope and at a much lower frequency, biasing the resulting parameters. In three stars, the estimated $\nu_{\mathrm{max}}$ used in the prior was significantly higher than the observed value, being overestimated by factors of three, five and two. In these cases, the parallaxes reported in the ATL were drawn from XHIP rather than GDR2. The $\nu_{\mathrm{max}}$ predicted using GDR2 parallaxes produced a prior more consistent with the measured $\nu_{\mathrm{max}}$. The remaining stars could be divided into two sets: Targets with less than one full sector of data and high-$\nu_{\mathrm{max}}$ targets with low mode amplitudes. The resulting low signal-to-noise could impact the determination of $\Delta\nu$. Values of $\nu_{\mathrm{max}}$ and $\Delta \nu$ for these outliers have been removed.

In Fig. \ref{fig:Seis HR} the stars in the ATL sample cluster about the base of the red giant branch and extend toward the main sequence. The peak of the distribution falls just above the \textit{Kepler} long cadence Nyquist frequency, populating the previously sparsely sampled region. The density falls off toward higher $\nu_{\mathrm{max}}$, which is likely a result of the decreasing mode amplitude.

\subsection{Large Sample}
Here, we report both $\nu_{\mathrm{max}}$ and $\Delta \nu$ for all but 62 stars. For these outliers, we found similar issues to those discussed in the ATL. Additionally, we noted 16 stars where the probability distributions in $\nu_{\mathrm{max}}$ were multi-modal. The remaining targets vastly outnumber those from the ATL, and span the red giant branch (as seen on Fig. \ref{fig:Seis HR}). The density increases with decreasing $\nu_{\mathrm{max}}$ until it peaks at $\approx 49 \mu$Hz. At the high-$\nu_{\mathrm{max}}$ tail, we note an overlap between the ATL and Large Sample. There are 119 stars that did not appear in the ATL, despite showing oscillations at frequencies above 240 $\mu$Hz. Of these, just under half lie near the ATL cutoff, with $240 < \nu_{\mathrm{max}} < 300 \mu$Hz. A total of 63 stars, however, are above 300 $\mu$Hz in a region that should be included in the ATL. There are several reasons why these targets could have been omitted. The estimate of $\nu_{\mathrm{max}}$ in the ATL was a function of $T_{\mathrm{eff}}$, such that a significant underestimate on the latter could have pushed the former beyond the enforced 240 $\mu$Hz cut. Calculating T$_{\mathrm{eff}}$ for the additional targets using the methods stated in the ATL, we did not find a systematic underestimate compared to GDR2, with values agreeing to within 10\% in over 90\% of the targets. The other possibility is that the detection probabilities for these targets were underestimated. This could be the result of an overestimated noise level caused by, for example, underestimating the size of the predicted pixel mask or a greater degree of contamination from background sources. A full comparison between the predictions in the ATL and the observed yield is reserved for future work.

\begin{figure}
    \centering
    \includegraphics[width=0.9\linewidth]{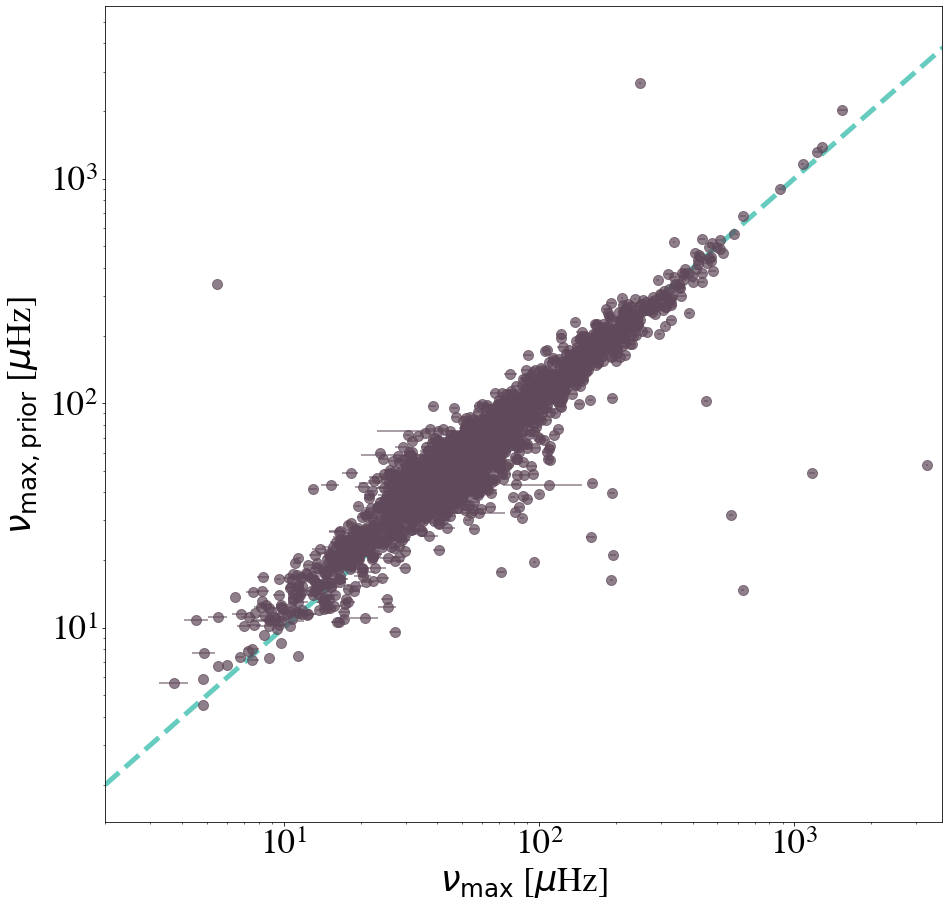}
    \caption{Prior $\nu_{\mathrm{max}}$ ($\nu_{\mathrm{max, prior}}$) versus measured values in short cadence targets. Blue dashed line represents the 1-1 line.}
    \label{fig:prior numax short}
\end{figure}

According to the main aim of the TESS mission \citep{2014SPIE.9143E..20R}, we expected targets proposed for 120-second cadence would be less evolved than the giants presented here. Although targets were also selected if they were brighter than $T_{\mathrm{mag}}$ = 6 (where T$_{\mathrm{mag}}$ is magnitude of the star for the TESS instrument response), which would preferentially select bright giants, we found that $\approx$ 80\% of the giants in which we detected solar-like oscillations were fainter than this limiting magnitude.  We therefore checked that the oscillations occur in the star associated with the TIC number being searched, rather than another star in the mask, by comparing the $\nu_{\mathrm{max}}$ value predicted by the prior to the detected value (see Fig. \ref{fig:prior numax short}). In general, the ratio of the two was close to unity, indicating the detected envelope belongs to the target in question. 

\begin{figure}
    \centering
    \includegraphics[width=1\linewidth]{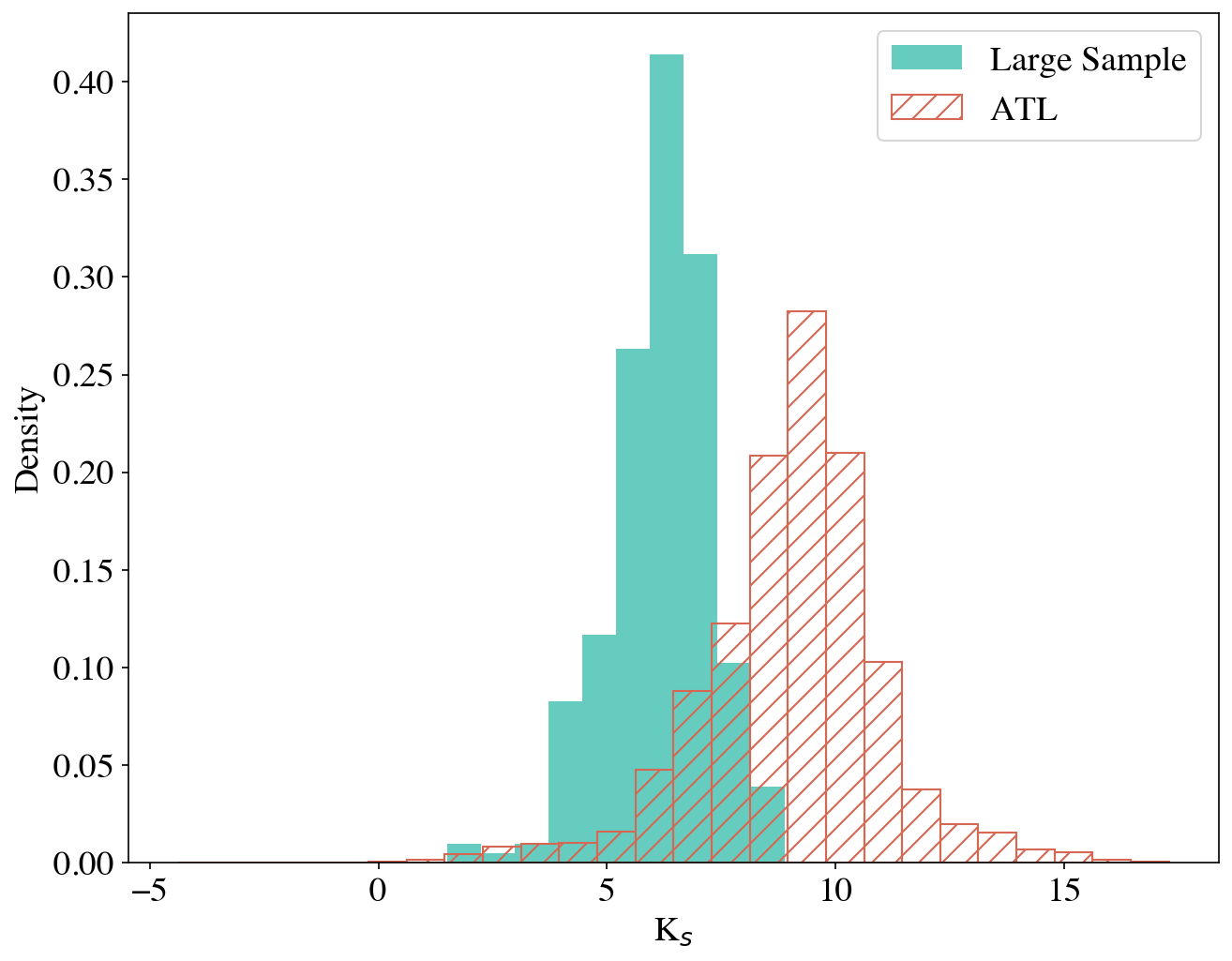}
    \caption{Distribution in $\Ks$-band magnitude for the detections from the Large Sample which have values of $\nu_{\mathrm{max}} > 240 \mu$Hz, and those made from targets included in the ATL.}
    \label{fig:numax short atl}
\end{figure}

\subsection{20-Second cadence}
\begin{figure}
    \centering
    \includegraphics[width=0.9\linewidth]{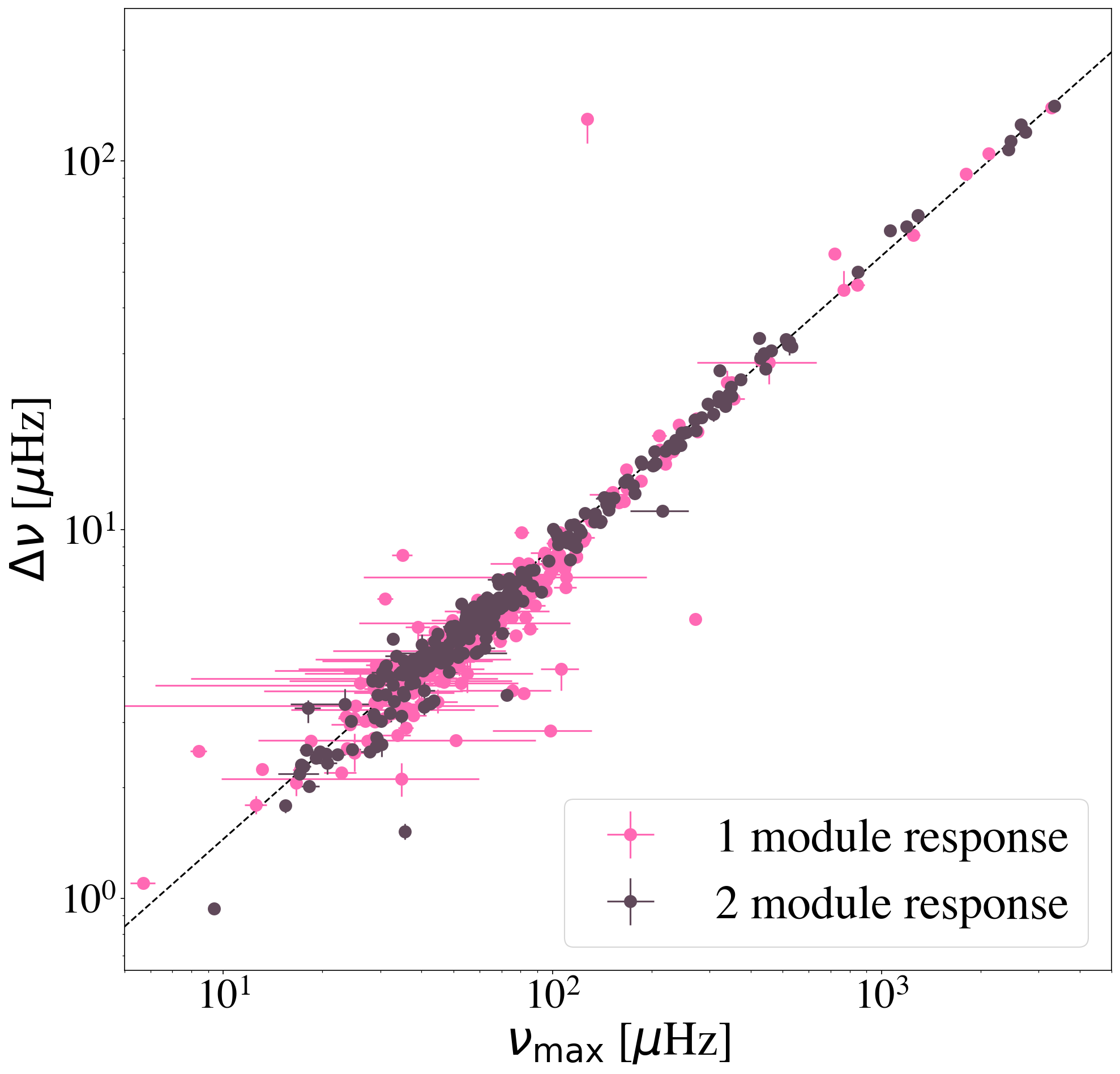}
    \caption{$\Delta \nu$ versus $\nu_{\mathrm{max}}$ for validated solar-like oscillators from the 20-second cadence set. Stars producing flags in only one module in pink, those producing flags in both are in brown.}
    \label{fig:dnu numax fast}
\end{figure}

The measured values of $\Delta \nu$ and $\nu_{\mathrm{max}}$ can be seen in Fig. \ref{fig:dnu numax fast}. A total of 16 stars were removed by manual identification. We note that  the uncertainties on $\Delta \nu$ presented in Fig. \ref{fig:dnu numax fast} appear larger than those presented in Fig. \ref{fig:dnu numax 733} (the ATL sample), which is likely due to the population of targets at $\nu_{\mathrm{max}}$ $<$ 100 $\mu$Hz. The mean fractional uncertainty on the measured $\Delta \nu$ in the 20-second cadence sample is 2.1\%, approximately consistent with 1.9\% in the ATL sample. Again, we find the population is dominated by evolved stars, with the distribution peaking at a $\nu_{\mathrm{max}}$ value of 50 $\mu$Hz.

We note the presence of a detection in an oscillator observed by $\textit{Kepler}$ (KIC 6106415; HD 177153; "Perky"), which is a clear outlier in Fig. \ref{fig:dnu numax fast}. The algorithm reports a $\Delta \nu$ of 131 $\mu$Hz despite measuring a $\nu_{\mathrm{max}}$ of 127 $\mu$Hz. Using \textit{Kepler} data oscillations were identified at $\nu_{\mathrm{max}}$ = 2249 $\mu$Hz \citep{2017ApJ...835..172L}. The envelope we detected at 127 $\mu$Hz (which can be visually identified) appears to be on another red giant in the pixel mask. Therefore the prior has caused an erroneous measurement of $\Delta \nu$.

\section{Conclusions}
Applying the algorithm introduced by \citet{Nielsen2021} to 120-second and 20-second cadence observations from the TESS mission spanning Sectors 1 to 46, we have detected solar-like oscillations in a total of 4177 targets. Of these, 12 belong to a set of previously reported solar-like oscillators, 752 to stars that appeared in the ATL and 486 were detected using 20-second cadence data. The remaining are targets brighter than 11th magnitude in 2MASS $K_{\mathrm{S}}$, with temperatures in the range 4500K < $\Teff$ < 6500K, observed in 120-second cadence. Since Sector 46, data for additional sectors have been released. We leave the analysis of these and the data collected in upcoming sectors as we approach the end of the first extended mission for future work.

All catalogued targets have been manually vetted to confirm the presence of oscillations. We note that signals from eclipsing binaries, classical pulsators, or transiting planetary bodies can cause false positive detections. Therefore, we highlight that when using the algorithm presented in N22 with very large data sets, the more conservative approach (asking for responses in both modules) is the most effective at reducing the amount of manual vetting required. 

We have extended the work of N22 to include methods to measure the global asteroseismic parameters, $\nu_{\mathrm{max}}$ and $\Delta \nu$. We introduced a new model to parameterize the collapsed ACF to produce a probability density for $\Delta \nu$. Applying this technique to the catalogue of detections, we measured the global asteroseismic parameters for 98\% of the targets. Overlaying these stars on the asteroseismic HR diagram ($\nu_{\mathrm{max}}$ and $\Teff$) allowed us to confirm the ATL successfully identified the least evolved stars, with little overlap in the remaining detections. The small set of stars that appear to have been missed by the ATL cluster about $K_{\mathrm{S}}$ = 6 mag, a region where the GDR2 astrometric solutions are known to have inferior astrometry \citep{2018A&A...616A...2L}, suggesting an issue in the parallaxes.
 
This catalogue has demonstrates the significant contribution the TESS mission can make to the field of asteroseismology. Isolating targets from the ATL, the increase in the number of detections between the 280 $\mu$Hz cut-off enforced by the 30-minute FFI observations and the upper edge of our catalogue at around 1000 $\mu$Hz is at least 2-fold on the detections made in \textit{Kepler} data. With the inclusion of the stars detected in 120-second cadence that did not appear in the ATL, we were able to use a homogeneous data set to measure asteroseismic values in solar-like oscillators from the subgiant regime through the red giant branch.

\begin{acknowledgements} 
Funding for the TESS mission is provided by the NASA's Science Mission Directorate. E.J.H., W.J.B. and G.R.D. acknowledge the support of Science and Technology Facilities Council. T.R.B. acknowledges support from the Australian Research Council (Discovery Project DP210103119). M.B.N. acknowledges support from the UK Space Agency. C.K. is supported by Erciyes University Scientific Research Projects Coordination Unit under grant number DOSAP MAP-2020-9749. D.B. acknowledges support from NASA through the Living With A Star Program (NNX16AB76G) and from the TESS GI Program under awards 80NSSC18K1585 and 80NSSC19K0385. D.H. acknowledges support from the Alfred P. Sloan Foundation and the National Aeronautics and Space Administration (80NSSC21K0652). T.S.M. acknowledges support from NASA grant 80NSSC22K0475. The authors acknowledge use of the Blue-BEAR HPC service at the University of Birmingham. This paper includes data collected by the Kepler mission and obtained from the MAST data archive at the Space Telescope Science Institute (STScI). Funding for the Kepler mission is provided by the NASA Science Mission Directorate. This work has made use of data from the European Space Agency (ESA) mission Gaia (\url{https://www.cosmos.esa.int/web/gaia}), processed by the Gaia Data Processing and Analysis Consortium (DPAC, \url{https://www.cosmos.esa.int/web/gaia/dpac/consortium}). Funding for the DPAC has been provided by national institutions, in particular the institutions participating in the Gaia Multilateral Agreement. This publication makes use of data products from the Two Micron All Sky Survey, which is a joint project of the University of Massachusetts and the Infrared Processing and Analysis Center/California Institute of Technology, funded by the National Aeronautics and Space Administration and the National Science Foundation. 
\end{acknowledgements}

\bibliographystyle{aa} 
\bibliography{test}

\appendix

\section{Additional Plots}\FloatBarrier

\begin{figure}
    \centering
    \includegraphics[width=1\linewidth]{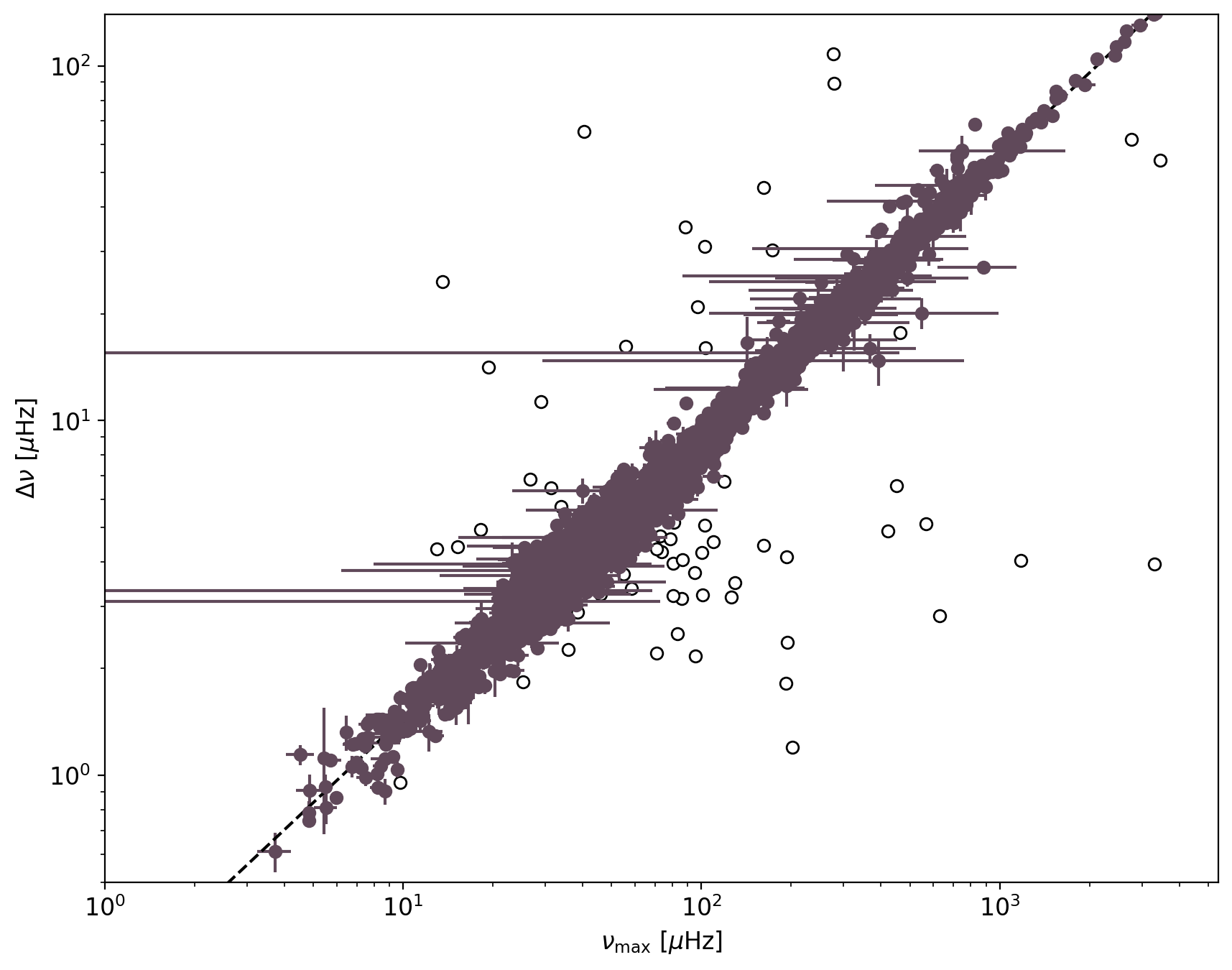}
    \caption{$\Delta \nu$ as a function of $\nu_{\mathrm{max}}$ for targets in the final catalogue in brown. Values that were removed after manual identification are shown as open circles.}
    \label{fig:bad globals}
\end{figure}

\begin{figure}
    \centering
    \includegraphics[width=0.9\linewidth]{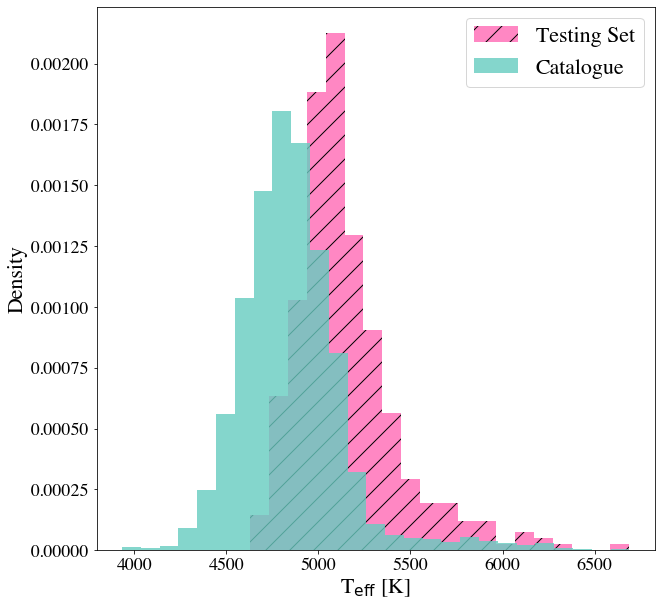}
    \caption{Distribution of T$_{\mathrm{eff}}$ for targets used in the testing set of 400 oscillators from N22 compared to those in the catalogue reported here.}
    \label{fig:Teff big test}
\end{figure}

\begin{figure}
    \centering
    \includegraphics[width=0.9\linewidth]{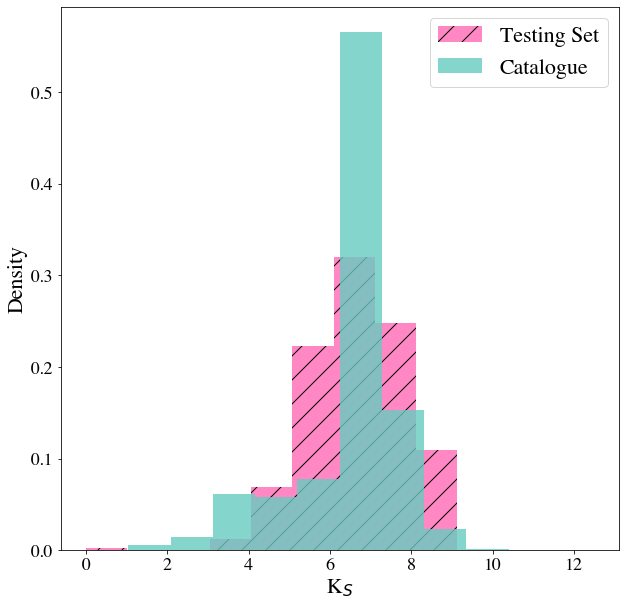}
    \caption{As in figure \ref{fig:Teff big test} with 2MASS K$_{\mathrm{S}}$ magnitude}
    \label{fig:Kmag big test}
\end{figure}

%
%

\end{document}